\documentclass[preprint]{aastex}
\usepackage{emulateapj5}
\usepackage{apjfonts}
\usepackage{natbib}

\input psfig.tex

\def\aa{{A\&A}}
\def\aas{{ A\&AS}}
\def\aj{{AJ}}
\def\al{$\alpha$}

\def\amin{$^\prime$}
\def\annrev{{ARA\&A}}
\def\apj{{ApJ}}
\def\apjs{{ApJS}}
\def\asec{$^{\prime\prime}$}

\def\cc{cm$^{-3}$}
\def\deg{$^{\circ}$}

\def\cc{cm$^{-3}$}
\def\e#1{$\times$10$^{#1}$}
\def\etal{{et al. }}
\def\fsec{\hbox{$.\mkern-4mu^s$}}

\def\flux{erg s$^{-1}$ cm$^{-2}$}

\def\hst{{\it HST}}
\def\kms{km s$^{-1}$}

\def\lax{{$\mathrel{\hbox{\rlap{\hbox{\lower4pt\hbox{$\sim$}}}\hbox{$<$}}}$}}
\def\gax{{$\mathrel{\hbox{\rlap{\hbox{\lower4pt\hbox{$\sim$}}}\hbox{$>$}}}$}}
\def\simlt{\lower.5ex\hbox{$\; \buildrel < \over \sim \;$}}
\def\simgt{\lower.5ex\hbox{$\; \buildrel > \over \sim \;$}}
\def\lum{erg s$^{-1}$}
\def\mbh{{$M_{\rm BH}$}}

\def\mnras{{MNRAS}}
\def\nat{{Nature}}
\def\pasp{{PASP}}

\def\percm2{cm$^{-2}$}
\def\peryr{yr$^{-1}$}

\def\solum{$L_\odot$}
\def\solmass{$M_\odot$}

\slugcomment{To appear in
{\it The Astrophysical Journal}.}
\lefthead{Ho, Terashima, \& Ulvestad}
\righthead{``Active'' Nucleus of M32}

\begin{document}

\title{Detection of the ``Active'' Nucleus of M32}

\author{Luis C. Ho\altaffilmark{1}, Yuichi Terashima\altaffilmark{2}, 
and James S. Ulvestad\altaffilmark{3}}

\altaffiltext{1}{The Observatories of the Carnegie Institution of Washington, 
813 Santa Barbara St., Pasadena, CA 91101.}

\altaffiltext{2}{Institute of Space and Astronautical Science, 3-1-1 Yoshinodai, Sagamihara, Kanagawa 229-8510, Japan.}

\altaffiltext{3}{National Radio Astronomy Observatory, P.O. Box O, Socorro, 
NM 87801.}

\begin{abstract}
M32 hosts a 2.5\e{6} \solmass\ central black hole, but signs of nuclear activity 
in this galaxy have long eluded detection.  We report the first conclusive 
detection of the nucleus of M32 in X-rays, based on high-resolution, sensitive 
observations made with {\it Chandra}.  The 2--10 keV luminosity is merely 
9.4\e{35} \lum, $\sim 3\times 10^{-9}$ of the Eddington luminosity of the 
black hole.  Weak diffuse emission, consistent with thermal gas at a 
temperature of 0.37 keV, is seen in an annular region $\sim$100 pc from the 
center.  We also present a deep, moderately high-resolution (0\farcs9), radio 
continuum observation obtained with the Very Large Array, which places a tight 
upper bound on the nuclear radio power at 8.4~GHz.  We combine these new 
measurements with upper limits at other wavelengths to discuss implications 
for the nature of accretion onto the central black hole.
\end{abstract}

\keywords{galaxies: active --- galaxies: individual (M32) --- galaxies: 
nuclei --- radio continuum: galaxies --- X-rays: galaxies}

\section{Introduction}

Along with our Galaxy and M31, M32 holds special significance as one 
of the three galaxies in the Local Group with a classical bulge component 
that also seems to host a massive black hole.  Since the initial work of Tonry 
(1984, 1987), the central stellar kinematics of M32 have been intensively 
scrutinized with increasingly elaborate data and methods of analysis (e.g., 
Dressler \& Richstone 1988; Richstone, Bower, \& Dressler 1990; van~der~Marel 
et al. 1994;  Qian et al. 1995;  Bender, Kormendy, \& Dehnen 1996; 
van~der~Marel et al. 1998; Joseph et al. 2001).  The latest efforts by Verolme 
et al. (2002) pinpoint the central dark mass in M32, presumed to be a black 
hole, to \mbh\ = $(2.5\pm0.5)\times 10^6$ \solmass.

The three Local Group galaxies with detected black holes serve as important 
laboratories for studying nuclear activity.  The Galactic Center source 
Sgr~A$^*$ and the nucleus of M31 both emit very weak, but detectable, 
nonstellar radiation in the radio and X-ray bands, considerations of which 
have led to new insights on accretion physics (Melia \& Falcke 2001, and 
references therein; Liu \& Melia 2001).  In this regard, M32 has remained 
stubbornly elusive.  All previous efforts to detect nonstellar emission from 
its nucleus have yielded either null or ambiguous results (Eskridge, White, \& 
Davis 1996; Loewenstein et al. 1998; Zang \& Meurs 1999).

Here we report the first conclusive detection of the nucleus of M32 in X-rays, 
based on high-resolution, sensitive observations made with {\it Chandra}.  We 
also present a deep radio continuum image obtained with the VLA, which 
places a tight upper limit at 8.4 GHz.  We combine these new 
measurements with upper limits at other wavelengths to discuss implications 
for the nature of accretion onto the central black hole.

Throughout this paper, we adopt a distance of 810 kpc for M32, based on the 
surface brightness fluctuation measurements of Tonry et al. (2001).

\vskip 0.3in
\section{Observations and Results}

\subsection{X-rays}

Our analysis is based on archival data acquired with {\it Chandra}\ (Weisskopf 
et al. 2002) using the Advanced CCD Imaging Spectrometer (ACIS).  The 
observations were performed on 2001 July 24 and 28.  We reprocessed the data 
with CIAO\footnote{http://asc.harvard.edu/ciao.} (version 2.2.1), using the 
calibration files CALDB 2.17.  In the first observation, the time 
intervals affected by background flares were discarded.  The background was 
high at the beginning of the second observation (first half about 3 times 
higher than the latter half), and gradually decreased. Because this does not 
affect the main results of our paper, we did not exclude the high background 
interval in the second observation in order to achieve better photon 
statistics.  The effective exposure times for the two observations are 
36.4 and 14.4 ks.

Figure~1 shows the combined {\it Chandra}\ image of the central 
30\asec$\times$30\asec\ (120 pc $\times$ 120 pc) region of M32, in the 0.3--8 
keV band.  Three unresolved X-ray sources, which we label X--1, X--2, and 
X--3, are clearly visible. Their positions, measured using the wavelet 
source-detection tool ``wavdetect,'' are summarized in Table 1.  The 90\%
astrometric error radius\footnote{http://asc.harvard.edu/cal/ASPECT/celmon.},
roughly symmetrical in right ascension and declination, is $\sim$0\farcs6.

The position of X--1 coincides with the 2MASS position of the nucleus of M32, 
as given in NED: $\alpha=00^h42^m$41\fsec830, $\delta$ = 
40\deg51\amin54\farcs50 (J2000).  The 1 $\sigma$ uncertainty of the 2MASS 
position is 0\farcs75 (Beichman et al. 1998), comparable to that of the 
{\it Chandra}\ position.  Based on this positional coincidence and the known 
reliability of the {\it Chandra}\ and 2MASS astrometry, we identify X--1 as 
the X-ray counterpart of the galaxy nucleus.  X--3 is the brightest source in 
the central region, and it lies 8\farcs4 (33 pc) to the southeast of X--1.  
Since it is much brighter than the other two sources in the field (see Table 
2), and its observed flux, 9.6\e{-13} \flux\ in the 2--10 keV band, is 
comparable to that previously measured with {\it ASCA}\ (Loewenstein et al. 
1998), the previously published {\it ASCA}\ and {\it ROSAT}\ fluxes

\begin{figure*}[t]
\centerline{\psfig{file=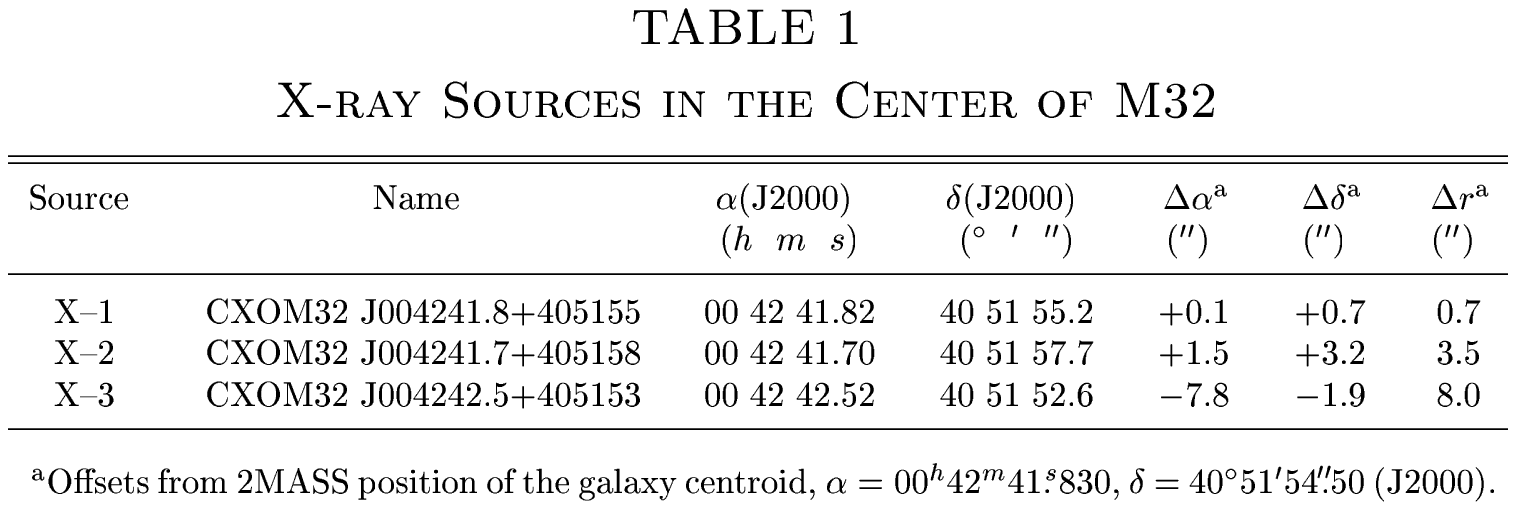,width=18.5cm,angle=0}}
\end{figure*}

\begin{figure*}[t]
\centerline{\psfig{file=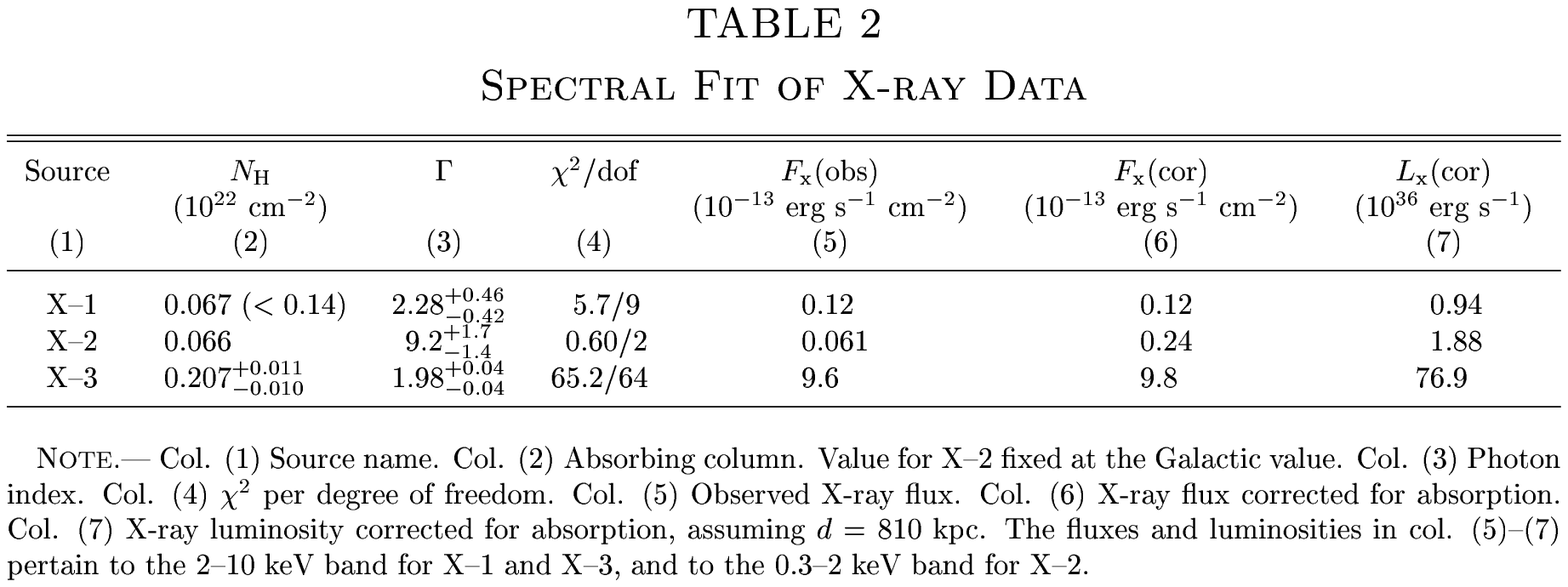,width=18.5cm,angle=0}}
\end{figure*}

\noindent
(Eskridge 
et al. 1996; Loewenstein et al. 1998) were probably dominated by X--3.  
Following the arguments given by Loewenstein et al. (1998), X--3 is most 
likely a low-mass X-ray binary. The faintest of the three sources, X--2, is 
separated from X--1 by 2\farcs4 (9.4 pc) to the northwest; it has a very soft 
spectrum and is 

\vskip 0.3cm

\psfig{file=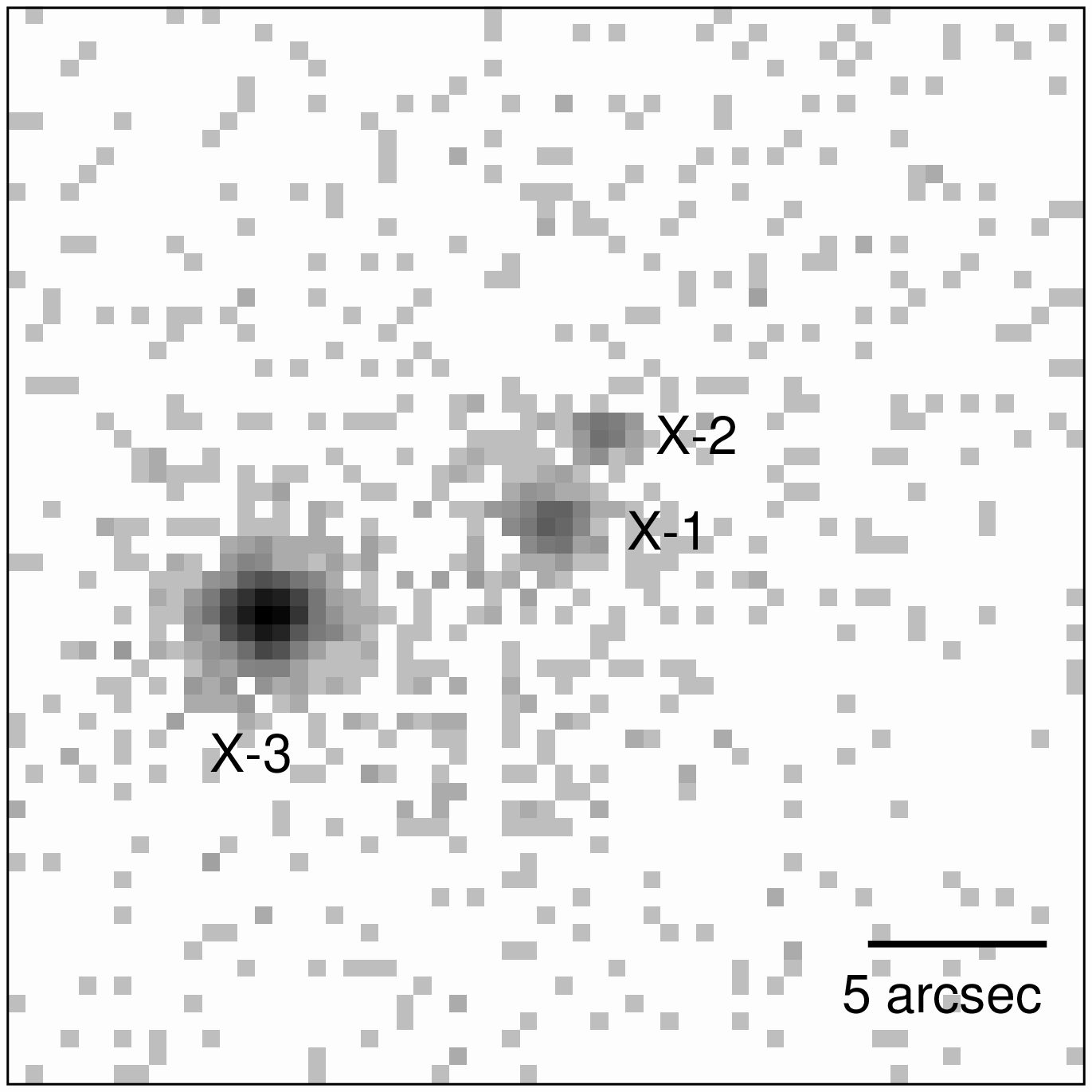,width=8.5cm,angle=0}
\figcaption[fig1.ps]{
{\it Chandra}\ image of the central 30\asec$\times$30\asec\ of the nucleus of
M32 in the 0.3--8 keV band.  The three X-ray sources are labeled.  North is 
up and East to the left.
\label{fig1}}
\vskip 0.3cm

\noindent
visible only at low energies (0.3--0.5 keV).

X-ray spectra were extracted for the three sources, after subtracting a 
background measured in the region immediately surrounding each source.  Source 
X--3 is relatively bright, and the effect of pile up, the tendency for two or 
more photons to be detected as a single event in bright sources, is 
nonnegligible; for this source, we applied the pile-up model in the 
spectral-fitting package XSPEC (version 11.2.0).  The spectra of the three 
sources can be well fitted by a simple model of a power law modified by 
absorption\footnote{The spectrum of source X--2 can also be fitted with a 
model of a blackbody with $kT\approx 40$ eV and fixing $N_{\rm H}$ to the 
Galactic value of 6.6\e{20} cm$^{-2}$ (Stark et al. 1992).  The fit has a 
$\chi^2 = 2.7$ for 2 degrees of freedom.}.  In all cases, the spectral 
parameters for the first and second observations are consistent with each 
other.  Table~2 summarizes the results of the spectral fits for the combined 
spectra of the two observations.  All the errors are at the 90\% confidence 
level for one parameter of interest ($\Delta \chi^2 = 2.7$).  The spectrum of 
the nucleus is shown in the left panel of Figure~2.

The light curves of the three sources show no significant short-term 
variability in either observation. The mean count rate of X--3 decreased 
slightly from the first to the second observation (0.11 counts s$^{-1}$ to
0.097 counts s$^{-1}$). No flux changes between the observations are seen for 
the other two sources.

To search for diffuse emission around the nuclear region, we made spectra of 
annular regions using the combined data of the two observations. The three 
nuclear point sources completely dominate the signal in the innermost region 
of the image ($r$ \lax\ 15\asec), making it impossible to place any useful 
constraints on the possible presence of a diffuse component.  The annular 
region between $r$ = 15\asec\ and 44\asec, however, {\it does}\ reveal 
statistically significant counts above the background, which we estimate 

\vskip 0.3cm
\begin{figure*}[t]
\centerline{\psfig{file=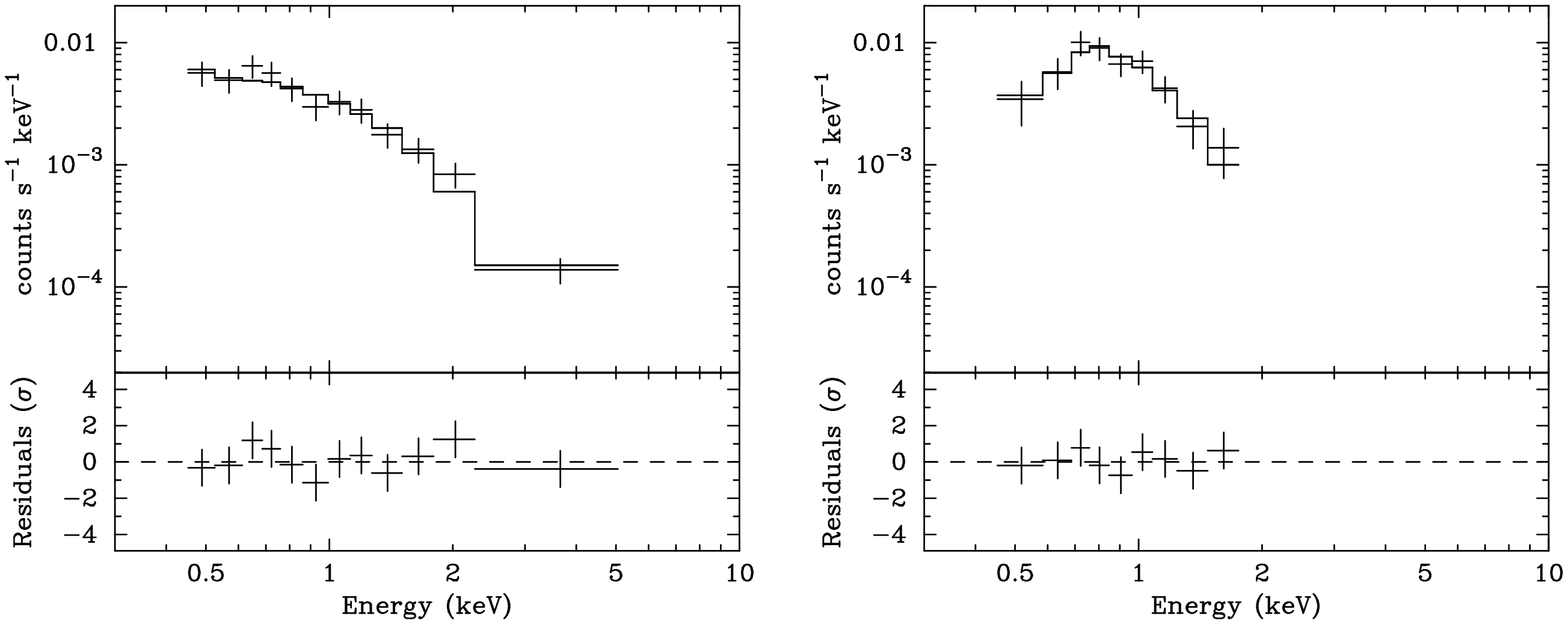,width=19.5cm,angle=270}}
\figcaption[fig2.ps]{
Spectrum of the nucleus ({\it left}) and of the diffuse emission in the
annular region 15\asec--44\asec\ centered on the nucleus ({\it right}).  The
best-fitting model for the nucleus is a power law with a photon index of
$\Gamma = 2.28^{+0.46}_{-0.42}$; for the diffuse emission, the best-fitting
model is that of a thermal plasma with a temperature of $kT$ = 0.37 keV.
The bottom panels show the residuals between the data and the model.
The vertical bars represent $\pm 1\,\sigma$ uncertainties.
\label{fig2}}
\end{figure*}
\vskip 0.3cm

\begin{figure*}[t]
\centerline{\psfig{file=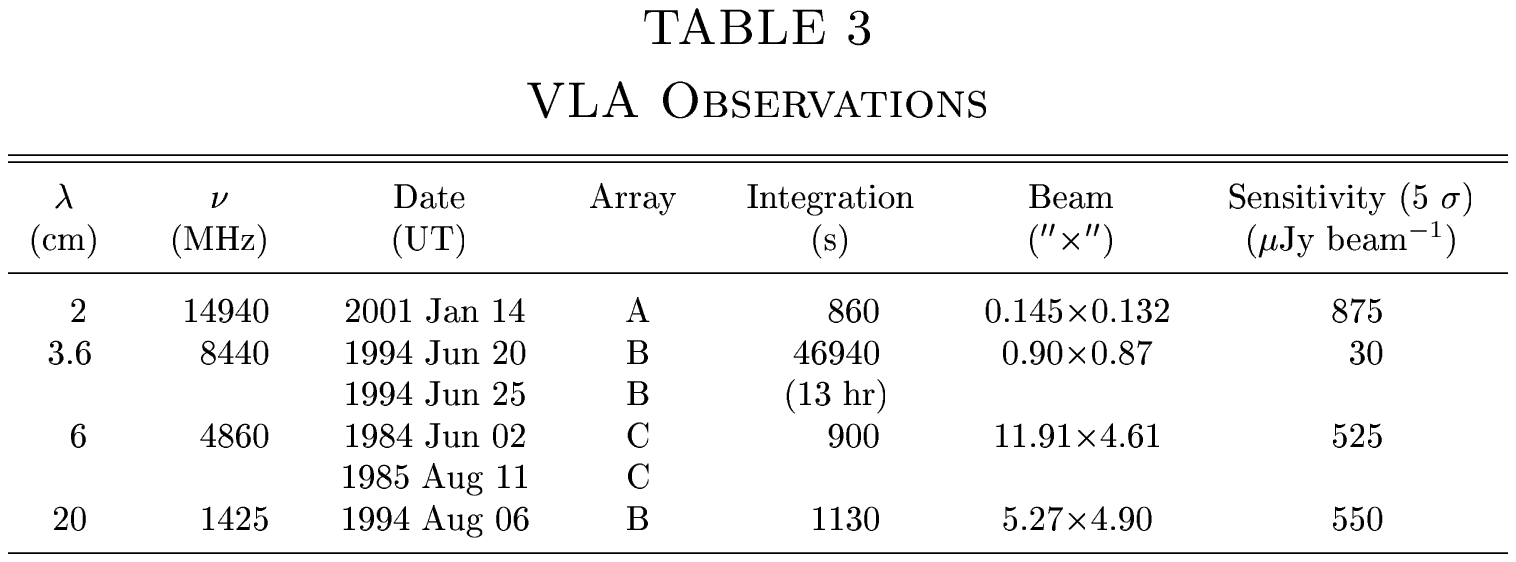,width=18.5cm,angle=0}}
\end{figure*}

\noindent
from 
the annulus centered at $r$ =  44\asec\--118\asec.  The spectrum of this 
component, displayed in the right panel of Figure~2, shows emission peaking 
around 0.7--0.8 keV, consistent with the Fe-L emission-line complex from a 
hot thermal gas.  

Figure~3 gives a visual representation of the spatial distribution of the 
extended emission in the energy band 0.5--2 keV. We use only the first 
observation in order to avoid the high background in the second observation.  
To construct the surface brightness profile, we calculated the background 
level from blank-sky observations at exactly the same locations on the CCD
chip where science data were extracted.  The blank-sky data, taken from the
{\it Chandra}\ data center, have an exposure time of 110 ks and are 
appropriate for a CCD temperature of --120$^{\circ}$ without correction for 
charge-transfer inefficiency. The background level is normalized using the 
count rates in the $r$ = 150--300 pixel (74\asec--148\asec) region.  In this 
plot, the bright source X--3 is located in the $r$ = 5\asec--10\asec\ bin.  
Extended emission in excess of the background is seen at $r$ \lax\ 50\asec, 
consistent with the spectral results. Note that the contribution from the 
wings of the point-spread function is negligible ($<$0.01\% of the peak) at 
5\asec\ away from a point source.

The soft emission seen in the inner annulus is most likely real and genuinely
associated with M32.  Neither the spectrum of the detector nor that of the
cosmic background shows such a spectral structure (Markevitch et al. 2003).  We
fitted the spectrum with a MEKAL thermal-plasma model (Liedahl, Osterheld, \&
Goldstein 1995) modified by absorption, assuming solar abundance ratios. The
spectrum extracted from the annulus $r$ = 44\asec--118\asec\ was used as
background.  The best fit, which has $\chi^2 = 2.2$ for 5 degrees of freedom, 
yields the following parameters (and their respective 90\% confidence range):
$N_{\rm H}=3.4 (1.5-5.2)\times10^{21}$ cm$^{-2}$, $kT = 0.37 (0.18-0.65)$ keV,
metallicity = 0.02 ($8\times10^{-4}-6\times10^{-2}$) solar, and volume 
emission measure = $2.9\times10^{60}$ cm$^{-3}$.  The flux in the 0.5--4 keV 
band, corrected for absorption, is $5.6\times10^{-14}$ erg s$^{-1}$ cm$^{-2}$; 
the corresponding luminosity is 4.4\e{36} \lum.  Assuming a volume filling 
factor of unity, the mean electron density is estimated to be $n_e$ = 0.069 \cc.

\vskip 0.3cm

\psfig{file=fig3.ps,width=8.5cm,angle=270}
\figcaption[fig3.ps]{
Surface brightness profile in the energy band 0.5--2 keV.  The filled circles
connected by the histogram are the data points for M32. The crosses
show the background estimated from blank-sky observations (see text).  The
counts in the $r$ = 5\asec--10\asec\ bin are dominated by the bright source
X--3.  Note the excess emission in the region $r\,\approx$ 15\asec--50\asec.
\label{fig3}}
\vskip 0.3cm

Note that the spectral parameters would be quite different had we fixed the 
column density to the Galactic value ($N_{\rm H}$ = 6.6\e{20} cm$^{-2}$).  
In this case, $kT = 1.0$ keV and the metallicity = 0.074 solar; however, the 
fit is statistically unacceptable ($\chi^2 = 9.5$ for 6 degrees of  freedom).

The metallicity of the diffuse gas formally obtained from our best-fitting 
model is markedly subsolar.  We note that this is not totally unexpected, 
since in elliptical galaxies the hot-gas metallicity correlates with both 
X-ray temperature and luminosity (Loewenstein et al. 1994; Davis 
\& White 1996; Matsushita, Ohashi, \& Makishima 2000).  M32's metallicity 
($\sim$0.02 solar) indeed roughly falls on these empirical relations (see 
Fig.~5 in Matsushita et al. 2000).

The temperature of the X-ray--emitting gas in elliptical galaxies scales with 
the central stellar velocity dispersion approximately as 
$kT \propto \sigma_*^{1.5}$, and generally the gas temperature is higher than 
the kinetic temperature of the stars (e.g., David \& White 1996).  M32, with 
$\sigma_* \approx 80$ \kms\ (Tonry 1984), follows these trends. Our new 
estimate of the X-ray gas temperature lies on the faint-end extrapolation of 
the $kT-\sigma_*$ relation of David \& White (1996).  The previous discrepancy 
noticed by David \& White (1996) is solely due to the contamination by 
the source X--3.

The above discussion explicitly assumes that the extended emission originates 
from hot gas.  The diffuse component in principle could come from a collection 
of unresolved discrete sources, but we consider this to be improbable.  X-ray 
binaries, the majority of which are of the low-mass variety, generally have 
harder spectra than that seen here (e.g., van Paradijs \& McClintock 1995).  
Individual stars do have soft spectra, but their luminosities are very low.

\subsection{Radio}

We retrieved several unpublished data sets from the archives of the Very Large 
Array (VLA){\footnote{The VLA is operated by the National Radio Astronomy 
Observatory, a facility of the National Science Foundation operated under 
cooperative agreement by Associated Universities, Inc.}}.  Observations are 
available at 1.4, 4.9, 8.4, and 15 GHz.  The data, summarized in Table~3, were 
edited, calibrated, imaged, and restored following routine procedures within 
AIPS (van~Moorsel, Kemball, \& Greisen 1996; for details, see, e.g., Ho \& 
Ulvestad 2001).  The images were made with natural weighting to maximize 
sensitivity, at the expense of a slight degradation in resolution.  By far the 
deepest image was that obtained at 8.4 GHz (3.6 cm), which resulted from a 
13~hr observation taken in the B configuration in 1994.  The image has a FWHM 
resolution (synthesized beam) of $\sim$0\farcs9 and an rms noise of $\sim6\,
\mu$Jy, which is close to the theoretical noise limit of the observation.  
None of the three {\it Chandra}\ sources are clearly detected.  There is a 
marginal (3 $\sigma$) excess that falls within the formal X-ray error 
circle\footnote{Note that the radio positions are accurate to much better than 
0\farcs1.} of the nuclear source, X--1, but there is signal at a comparable 
level of significance throughout the image.  Thus, we conservatively assign a 
5 $\sigma$ upper limit of 30 $\mu$Jy at 8.4 GHz; upper limits for the other 
frequencies are given in Table~3.  To our knowledge, the only previously 
published observation of M32 with sufficient resolution to place meaningful 
constraints on its nuclear radio emission is that by Heckman, Balick, \& Crane 
(1980); they list an upper limit of 1 mJy at 5 GHz and 6 mJy at 1.4 GHz.

The 8.4 GHz map reveals only one solid detection above 5 $\sigma$, with a flux 
density of $166\pm10\, \mu$Jy at $\alpha$ = $00^h42^m$45\fsec705, $\delta$ =
40\deg52\amin43\farcs77 (J2000), but it is located \gax 1\amin\ from the 
nucleus of M32.  It does not have an obvious X-ray counterpart in the 
{\it Chandra}\ image.

\vskip 0.3cm
\subsection{Ultraviolet, Optical, and Infrared}

A number of authors have studied the central region of M32 at high resolution
($\sim$0\farcs1) using the {\it Hubble Space Telescope (HST)}.  Table~4 
assembles the data pertinent to the unresolved component of the nonstellar 
nucleus.  None of the measurements have yielded a positive detection, but 
\hst\ does provide very tight upper limits.  The central light distribution of
M32 is extremely cuspy at optical (Lauer et al. 1998) and near-infrared 
(Ravindranath et al. 2001) wavelengths.  Our limits on the nonstellar 
contribution in the $V$ and $H$ bands come from detailed analysis of the 
two-dimensional light distribution of the central region of the galaxy 
(Ravindranath et al. 2001; Peng et al.  2002).  We also include the 10 $\mu$m 
limit published by Grossan et al. (2001).

\section{Discussion}

\subsection{M32 in Context}

Most, perhaps all, nearby galaxies with a bulge contain central massive black 
holes (Kormendy \& Gebhardt 2001), but, for reasons that are not yet well 
understood, many black holes are puzzlingly inactive.  The denizens of
the Local Group offer a good illustration.  The Galactic Center is known to 
host a $2.6\times 10^6$ \solmass\ black hole (Eckart \& Genzel 1997; Ghez et 
al. 1998) that is extremely dormant.  Apart from the radio band, the central 
source Sgr~A$^*$ has been convincingly detected only in the X-rays.  The 
2--10 keV X-ray luminosity in its normal quiescent state is merely $L_{\rm X} 
\approx 2\times 10^{33}$ \lum, flaring to $\sim 1\times 10^{35}$ \lum\ during 
brief periods (Baganoff et al. 2001).  The X-ray luminosity thus accounts for 
only $7\times 10^{-12}-3\times 10^{-10}$ of the Eddington luminosity.  The 
radio band contributes somewhat less, but still an energetically significant 
amount compared to the X-rays.  The time-averaged flux density of Sgr~A$^*$ 
at 8.4 GHz is $\sim$ 0.8 Jy (Zhao et al. 1992), which corresponds to a 

\begin{figure*}[t]
\centerline{\psfig{file=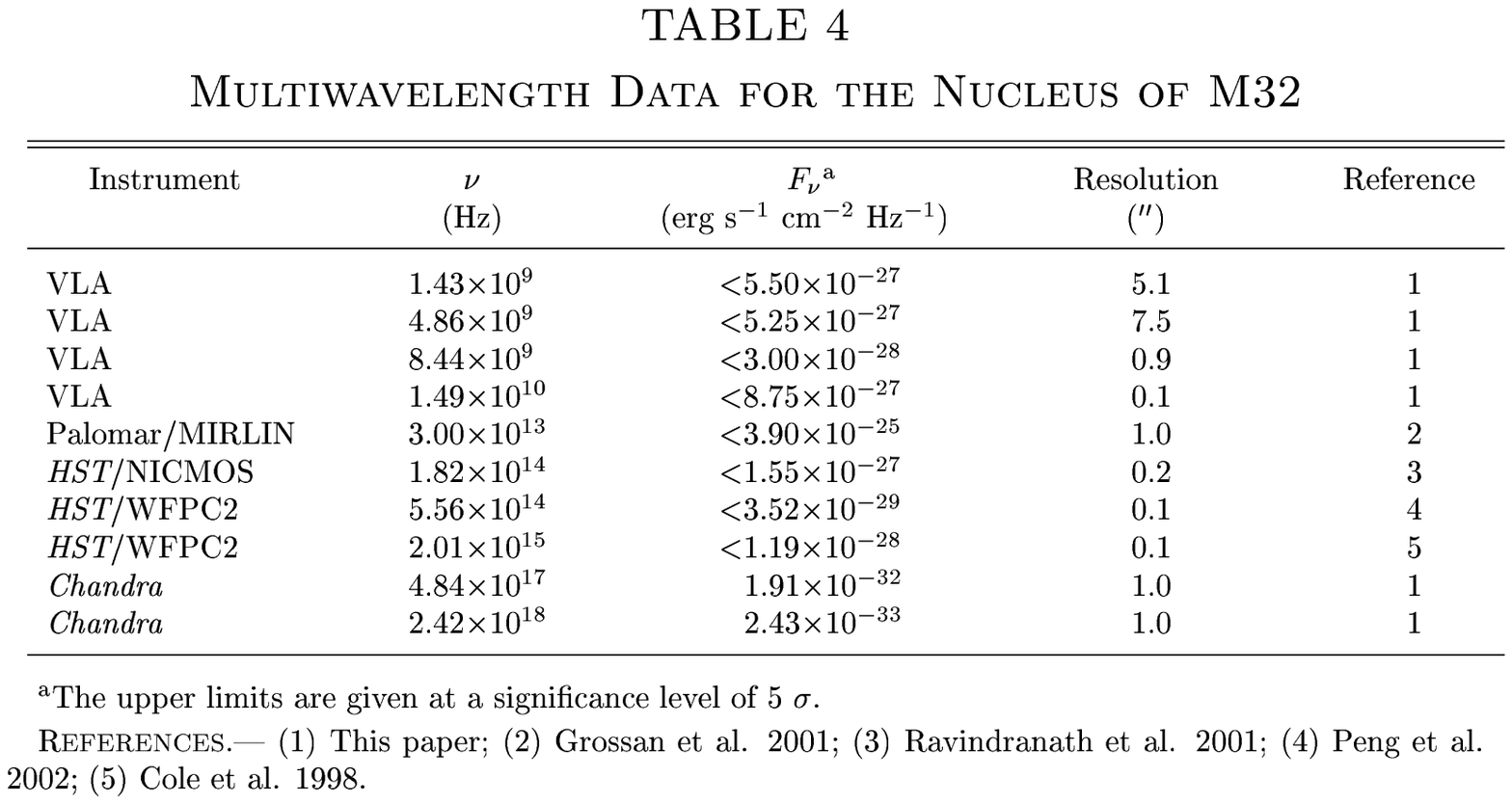,width=18.5cm,angle=0}}
\end{figure*}

\noindent
spectral luminosity of $L_{\rm R} \equiv \nu L_{\nu} \approx 6 \times 10^{32}$ 
\lum\ for an assumed distance of 8.5 kpc.  The ratio of radio to X-ray 
luminosity is log~($L_{\rm R}/L_{\rm X})\, \approx$ $-0.5$ and $-$2.2 in its 
quiescent and flaring state, respectively.

The nucleus of M31 is less extreme, but only moderately so.  Garcia et al. 
(2000) report $L_{\rm X} = 3.9\times 10^{36}$ \lum.  For \mbh\ = 
$4.5\times 10^7$ \solmass\ (Kormendy \& Bender 1999; Bacon et al. 2001), 
$L_{\rm X}/L_{\rm Edd} = 7\times 10^{-10}$.  M31 similarly has a central radio 
core, with a luminosity approximately one-third of that of Sgr~A$^*$ [flux 
density at 8.4 GHz $\sim$30 $\mu$Jy (Crane, Dickel, \& Cowan 1992), or 
$L_{\rm R} \approx 2 \times 10^{32}$ \lum\ for an assumed distance of 
800 kpc].  Unlike Sgr~A$^*$, however, here the relative fraction of radio to 
X-ray luminosity is significantly lower, log~($L_{\rm R}/L_{\rm X}) \approx 
-4.3$.

M32 is the third galaxy in the Local Group known to possess a massive black 
hole; its mass, \mbh\ = $(2.5\pm0.5)\times 10^6$ \solmass\ (Verolme et al. 
2002), is essentially identical to that of the Galactic Center.  As mentioned 
in the Introduction, all previous studies have failed to detect emission 
unambiguously associated with accretion onto the central black hole.  Eskridge 
et al. (1996) detected X-ray emission from the central region of M32, but the 
poor positional accuracy (9\asec) and low resolution (27\asec) of the 
{\it ROSAT}\ PSPC data precluded a definitive identification of the nucleus.  
Loewenstein et al. (1998) subsequently obtained higher resolution (5\asec) 
observations using the {\it ROSAT}\ HRI, which led them to suggest that most 
of the X-ray emission lies slightly offset from the optical position of the 
nucleus.  They attributed the X-ray emission to a single low-mass X-ray 
binary.  This conclusion, however, was later questioned by Zang \& Meurs 
(1999), who maintained that a nuclear origin for the X-ray emission cannot 
be ruled out.

The {\it Chandra}\ image presented in this paper, with its superior angular 
resolution (1\asec), positional accuracy (0\farcs6), and sensitivity resolves 
this longstanding debate.  Our analysis is also greatly aided by the recent 
availability of a more reliable position for the galaxy centroid from 2MASS.  
We find that the X-ray emission from the center of M32 arises not from one, 
but {\it three}\ discrete sources.  We propose that X--1 marks the ``active'' 
nucleus of M32.

With a 2--10 keV luminosity of 9.4\e{35} \lum, M32 formally posesses the 
second lowest luminosity X-ray nucleus known, slightly lagging behind even 
M31.  It is less than a factor of ten more luminous than Sgr~A$^*$ in its high 
state.  Relative to the Eddington rate, however, the luminous output of M32's 
nucleus ranks highest among the three Local Group galaxies, although it is 
still very sub-Eddington ($L_{\rm X}/L_{\rm Edd} = 3\times 10^{-9}$).  
Within the limited photon statistics, the X-ray spectrum is consistent with 
a simple power law, characterized by a photon index $\Gamma = 
2.28^{+0.46}_{-0.42}$.  The absorption column is low, $N_{\rm H} = 
6.7\times10^{20}$ cm$^{-2}$, comparable to the Galactic value along the line 
of sight to M32.  The spectral slope is essentially identical to that of 
Sgr~A$^*$ in its quiescent state ($\Gamma = 2.2^{+0.5}_{-0.7}$; Baganoff et 
al. 2001) and, within the errors, statistically similar to that of typical 
low-luminosity AGNs (Terashima et al. 2002), but much harder than the 
unusually soft spectrum found for the nucleus of M31 ($\Gamma = 4.5\pm1.5$; 
Garcia et al. 2000).

The radio upper limit at 8.4 GHz, 30 $\mu$Jy, translates into 
log~($L_{\rm R}/L_{\rm X}) < -3.7$.  This ratio is consistent with that 
observed in M31.  If the physical mechanism giving rise to the radio and X-ray 
emission is the same in the two galaxies, it suggests that M32's nucleus 
should be detectable with radio observations deeper by a factor of a few.  
This extrapolation, however, is clearly uncertain, in view of the very 
different X-ray spectral slopes of the two objects.  Moreover, it is difficult 
to predict what the relative strength of radio and X-ray emission ought to be, 
for low-luminosity nuclei evidently show a wide dispersion in their observed 
radio/X-ray luminosity ratios.  As noted above, in the Galactic Center 
log~($L_{\rm R}/L_{\rm X})\, \approx$ $-0.5$ to $-$2.2, whereas 
log~($L_{\rm R}/L_{\rm X})\, =$ $-$3.6, $-$2.9, and $-$3.4 for NGC 3147, 
NGC 4203, and NGC 4579, respectively (Ulvestad \& Ho 2001).  In the sample
studied by Terashima \& Wilson (2003), log~($L_{\rm R}/L_{\rm X}$) ranges from 
$-5$ to $-2$.

\vskip 0.6cm
\subsection{Constraints on the Nature of Accretion}

High-resolution images of M32 are available at ultraviolet, optical, 
near-infrared, and mid-infrared wavelengths, but the nucleus was undetected in 
any of those bands (Table 4).  Thus, we are unable to construct a meaningful
spectral energy distribution (SED) for the nucleus.  Nevertheless, we can 
utilize the X-ray measurement to estimate the total bolometric luminosity, 
under the assumption that the overall true SED of M32 is not too dissimilar 
from those of other low-luminosity AGNs (Ho 1999, 2002).  This assumption is 
not inconsistent with the data given in Table 4.  In low-luminosity AGNs, the 
X-ray power in the 2--10 keV band constitutes roughly 15\% of the bolometric 
luminosity (Ho 1999; Ho et al. 2000). Hence, in M32 $L_{\rm bol} \approx 
6\times10^{36}$ \lum, and $L_{\rm bol}/L_{\rm Edd} \approx 2\times 10^{-8}$. 

The exceedingly low nuclear accretion luminosity in M32 suggests that 
presently it must be largely devoid of fuel for accretion.  This indeed seems 
to be substantiated by observations.  Sensitive searches for 
molecular and atomic hydrogen have failed to detect any cold gas, down to 
a stringent upper limit of 11,000 \solmass\ in the inner $\sim$100 pc 
radius (Sage et al. 1998; Welch \& Sage 2001).  The spectra of Ho, Filippenko, 
\& Sargent (1997) show no trace of H\al\ emission to an equivalent-width limit 
of $\sim$0.25 \AA\ within an aperture of  2\asec$\times$4\asec\ (8 pc $\times$ 
16 pc), which corresponds to an H\al\ luminosity of $L_{{\rm H}\alpha}$ \lax\
2\e{36} \lum\ (Ho, Filippenko, \& Sargent 2003).  The mass of ionized hydrogen 
in an ionization-bounded nebula can be expressed as $M_{\rm H~II} = 
3.2\times10^{-33} \left(L_{{\rm H}\alpha}/n_e\right)$ \solmass, where $n_e$ 
is the electron density (e.g., Osterbrock 1989).  For an assumed value of 
$n_e$ = 100 cm$^{-3}$, not unreasonable for most galactic nuclei (Ho et al. 
2003), the mass in warm ($10^4$ K), ionized gas is $M_{\rm H~II}$ \lax\ 60 
\solmass\ --- vanishingly small.  The only direct evidence for gas in the 
vicinity of M32's nucleus comes from the {\it Chandra}\ data.  We detected 
diffuse, thermal plasma with a temperature of 0.37 keV in an annular region 
15\asec--44\asec\ ($\sim$60--170 pc) from the center.  It is unclear, however, 
whether this diffuse gas extends further in toward the nucleus.

Despite the lack of definitive evidence for the presence of gas in the 
innermost regions of M32, some gas {\it must}\ be there, namely that shed 
through normal mass loss by evolved stars. M32 has an extremely cuspy central
light distribution. The stellar density rises steeply toward the center as 
$\rho \propto r^{-1.5}$, reaching a luminosity density of 
$\sim 10^7\,L_{\odot, V}\,{\rm pc}^{-3}$ at $r = 0.1$ pc (Lauer et al. 1998).  
Now, the mass loss rate of an integrated population of evolved stars has been 
well characterized in a number of studies.  According to the calculations of 
Padovani \& Matteucci (1993), a stellar population following a Salpeter 
initial mass function with a lower-mass cutoff of 0.1 \solmass, an upper-mass 
cutoff of 100 \solmass, solar metallicity, and an age of 15 Gyr generates a 
mass loss rate of 
$\dot M_*\,\approx\,3\times10^{-11} \, \left(L/L_{\odot, V}\right)$ 
 $M_{\odot}\,{\rm yr}^{-1}$.  This value appears to be quite robust.  To 
within a factor of $\sim 2$, it agrees with other estimates based on similar, 
but not identical, assumptions (Faber \& Gallagher 1976; Jungwiert, Combes, \& 
Palous 2001; Athey et al. 2002).   Hence, within $r = 0.1$ pc 
$L \approx 4\times 10^4 \, L_{\odot, V}$, or 
$\dot M_*\,\approx\,1\times10^{-6}$ \solmass\ \peryr.  

In addition to stellar mass loss, we can also consider Bondi (1952) accretion 
of the hot gas, under the assumption that the diffuse component continues to 
extend inward toward the nucleus.  For a central object of mass $M$ permeated 
by gas with a density $n$ and temperature $T$, the Bondi accretion rate 
$\dot M_{\rm B} \propto M^2 n T^{-3/2}$.  Using parameters appropriate for 
M32, we find $\dot M_{\rm B} \approx 3\times 10^{-7}$ \solmass\ \peryr.  
This estimate makes the conservative assumption that the density profile 
remains flat at $r\,<$ 15\asec.  More likely, $n$ will rise toward the 
center, in which case $\dot M_{\rm B}$ will increase.  In any case, it 
appears that, to first order, $\dot M_* \approx \dot M_{\rm B}$.

Either rate, $\dot M_*$ or $\dot M_{\rm B}$, is tiny, but nonnegligible in 
terms of accretion luminosity, provided that it can be efficiently converted 
to radiation.  For example, the luminosity produced by a canonical optically 
thick, geometrically thin disk is given by $L_{\rm acc} = \eta \dot M c^2 = 
5.7\times10^{45} \left(\eta/0.1\right) \left(\dot M/M_{\odot}\, 
{\rm yr}^{-1}\right)$ \lum, where $\eta$ is the radiative efficiency.  For 
$\dot M$ = $\dot M_*$ as given above, $L_{\rm acc} = 6\times10^{39}$ \lum.
Clearly, this estimate severely violates, by about 3 orders of magnitude, 
the observed limits on the bolometric luminosity of the nucleus.

The above simple considerations lead to three possible implications.  First, 
our estimate of $\dot M_*$ or $\dot M_{\rm B}$ could be grossly in error.  
Although the Bondi rate is somewhat uncertain because of the lack of direct 
information on the gas density and temperature close to the nucleus, we 
believe that the stellar mass loss rate should be quite reliable because the 
central light distribution of M32 is very well measured (Lauer et al. 1998), 
because the stars in M32's nucleus are known to be evolved (e.g., Davidge 
1990), and because the prescription we adopted for computing $\dot M_*$ is 
known to be quite robust.  Second, if $\dot M_*$ is correct and 
$\dot M \approx \dot M_*$, the radiative efficiency must be very small; 
instead of 0.1, $\eta$ may be as low as $\sim 10^{-4}$.  This may not be 
surprising, in view of the extraordinarily low luminosity and accretion rate 
in M32's nucleus; under such conditions, its accretion flow is 
likely to be radiatively inefficient (see reviews by Narayan, Mahadevan,
\& Quataert 1998 and Quataert 2001).  And third, $\dot M_*$ could be correct 
but the true accretion rate may be much less than $\dot M_*$.  This would be 
the case if the majority of the gas gets ejected from the nucleus, for 
instance as a result of supernova explosions.  Alternatively, it is possible 
that $\dot M \ll \dot M_*$ because much of the gas in the accretion flow 
actually becomes unbound, as in the adiabatic inflow-outflow solution 
discussed by Blandford \& Begelman (1999).  

We consider gas removal by supernovae to be implausible in the case of M32. 
Its nucleus shows no trace of recent star formation, which effectively rules 
out core-collapse supernovae.  The expected rate of Type~Ia supernovae, on the 
other hand, is far too low to matter, as the following crude estimate 
indicates.  In elliptical galaxies, Cappellaro et al. (1997) determine the 
Type~Ia supernova rate to be 0.13 per century per $10^{10}\,L_{\odot}$ in the 
$B$ band.  Taking the $V$-band luminosity within $r = 0.1$ pc, $4\times 10^4$ 
\solum\ (\S~3.2), and a typical color of $B-V \approx 1$ mag for elliptical 
galaxies (e.g., Fukugita, Shimasaku, \& Ichikawa 1995), we estimate the 
Type~Ia supernova rate within the nucleus of M32 to be $2\times10^{-9}$ 
\peryr.  Assuming that each supernova liberates $10^{51}$ ergs, the 
time-averaged energy injection rate is only $6\times 10^{34}$ \lum, or 
$\sim$1\% of the X-ray luminosity of the diffuse gas.

\vskip 0.3cm

\section{Summary}

A sensitive, high-resolution {\it Chandra}\ image has firmly detected, for the 
first time, hard X-ray emission associated with the nucleus of M32.  The 
pointlike emission has a hard power-law spectrum and presumably arises from 
accretion onto its central massive black hole.  The signal is very faint: the 
2--10 keV luminosity is only 9.4\e{35} \lum, one of the lowest ever measured 
in a galactic nucleus.  We analyzed a deep 8.4~GHz map, but no radio 
counterpart of the X-ray core could be seen to a 5 $\sigma$ limit of 30 
$\mu$Jy.  The nucleus is also undetected in high-resolution images taken at 
ultraviolet, optical, near-infrared, and mid-infrared wavelengths.  These 
measurements allow us to constrain the bolometric luminosity 
of the nucleus, which we estimate to be only $\sim 2\times 10^{-8}$ of the 
Eddington luminosity of the black hole.  The quiescence of the nucleus of 
M32 is not unexpected, given the present gas-poor conditions prevalent in the 
circumnuclear regions of the galaxy.  Nevertheless, we argue that M32's dense
nuclear cusp ought to supply sufficient fuel through stellar mass loss to 
sustain a much more pronounced level of nuclear activity.   We discuss the 
consequences of this paradox, which implies that either the central accretion 
flow is extremely radiatively inefficient or that most of the gas escapes 
accretion.

\acknowledgements
The research of L.~C.~H. is funded by the Carnegie Institution of Washington 
and by NASA grants from the Space Telescope Science Institute (operated by 
AURA, Inc., under NASA contract NAS5-26555).  Y.~T. is supported by the 
Japan Society for the Promotion of Science.  This work has used the NASA/IPAC
Extragalactic Database (NED) which is operated by the Jet Propulsion
Laboratory, California Institute of Technology, under contract with NASA.
We thank Stephen Helsdon for help with the initial examination of the 
{\it Chandra}\ data and Chien Peng for reanalysis of the \hst\ $V$-band image.
We appreciate the constructive comments provided by an anonymous referee, 
who motivated us to evaluate the effect of Type~Ia supernovae discussed in 
\S~3.2.

%

%

\end{document}